# Southern Hemisphere Mid-latitude Atmospheric Variability of the NCEP-NCAR and ECMWF Reanalyses

Alessandro dell'Aquila[(1)], Paolo M. Ruti[(1)], Sandro Calmanti[(1)] and Valerio Lucarini[(2)]

[1]*Progetto Speciale Clima Globale, Ente Nazionale per le Nuove Tecnologie, l'Energia e l'Ambiente, Roma, Italy*

[2] *PASEF- Physics and Applied Statistics of Earth Fluids, Dipartimento di Matematica ed Informatica, Università di Camerino, Camerino (MC), Italy*

Corresponding author:

Dr. Alessandro Dell'Aquila

ENEA, C.R. Casaccia,

Via Anguillarese 301, 00060 S. Maria di Galeria, Rome, Italy

E-mail: alessandro.dellaquila@casaccia.enea.it

Phone:+39-06-30484072

Fax:+39-06-30484264

# Abstract

We compare the representation of the southern hemisphere midlatitude winter variability in the NCEP-NCAR and ERA40 reanalyses by adopting the Hayashi spectral technique. We find relevant discrepancies in the description of the atmospheric waves at different spatial and temporal scales. ERA40 is generally characterised by a larger variance, especially in the high frequency spectral region. In the southern hemisphere, even in the satellite period, the assimilated data are relatively scarce, predominately over the oceans, and they provide a weak constraint to the assimilation system. In the pre-satellite period the discrepancies between the two reanalyses are large and randomly distributed while after 1979 the discrepancies are systematic. This study suggests that, as for the winter mid-latitude variability in the southern hemisphere, a well-defined picture to be used in the evaluation of the realism of climate models is still lacking because of the non-consistency of the reanalyses.

# 1. Introduction

In the Southern Hemisphere (SH), the absence of a pattern of mountain chains that locks the phase of planetary scale waves implies that most part of the atmospheric variability is accounted for by eastward propagating waves trains at both high and low frequencies (James, 1994).

High frequencies roughly correspond to periods shorter than 10 days .On such time scale, the midlatitude atmospheric flow is dominated in both hemispheres by the growth and decay of baroclinic disturbances, which convert available potential energy into kinetic energy (Holton, 1992). In the SH, the baroclinic activity is stronger than in the NH and peaks around 50°S. Regionally, the high frequency disturbances are strongest in the southern Indian Ocean (Trenberth, 1991; Frederiksen and Frederiksen, 1993; Cuff and Cai, 1995), where the meridional heat fluxes maximise.

On the low frequency time-scale (10-50 days), the eastward propagating waves trains feature a characteristic spatial patterns, often referred to as Pacific South American (PSA), characterised by ultra long spatial scales (Robertson and Mechoso, 2002). A standing pattern associated to the spatial zonal wavenumber 3 and often related to blocking events is also present (Trenberth and Mo 1985, Raphael 2004).

Recently, a great interest has raised regarding the skill of the two reanalyses (Bromwich and Fogt 2004). Our aim is to analyze quantitavely the differences in the mid-latitude SH winters observed in the NCEP and ERA40 reanalyses in describing intraseasonal variability at different time and spatial scales. For this purpose, we adopt a spectral analysis technique recently introduced by Hayashi (1971, 1979) employed in the companion comparison study performed on the NH (Dell'Aquila et al., 2005 – D05 henceforth), where the winter variability has been analyzed in terms of standing and propagating waves. Morever, we employ measures of the bulk spectral properties of the waves, which have proved to be metrics of relevance in defining benchmarks for the assessment of the reliability of climate models (Lucarini et al., 2006).

The intercomparison of the description provided by the reanalyses for the SH atmospheric variability constitutes an appealing issue for several different reasons.

- Fewer observational data are available for the assimilation system in the SH. Unlike in the NH, the scarce amount of rawinsondes data does not provide a strong constraint to the observed three-dimensional thermal structure of the tropospheric disturbances in the SH. Here, the most relevant data source is the remote-sensing observational system, which represents a weaker constraint to the atmospheric vertical structure.
- The SH mid-latitudes troposphere is strongly coupled to the ocean and to the stratosphere..
- The absence of a relevant topographic forcing rules out the possibility that the observed discrepancies between the reanalyses can be attributed to significant differences in the interaction of the atmospheric flow with topography.
- All of these issues bear relevance for the initialization of service and research weather forecast models operating at various spatial scales.

Another study focusing on the reanalyses intercomparison in the SH (Hoskins and Hodges, 2002) employs a tracking method resulting in a detailed description of the geometry of the synoptic systems. Our spectral approach, following several studies in the '80s (Mechoso and Hartmann, 1982; Fraedrich and Kietzig, 1983; Hansen et al., 1989), complements the analysis of Hoskins and Hodges (2002) by featuring a more detailed description of the wave kinematics, while the information about their geometry is essentially lost.

The paper is organized as follows. In Section 2 we review the main differences between the reanalysis systems, describe the considered datasets, and present the Hayashi spectral technique. In Section 3 we show the results of Hayashi spectral analysis for the two datasets and discuss the emerging differences. In Section 4 we compare the spatial distribution of the variability for the two datasets before and after the introduction of satellites observations. In Section 5 we summarize the results obtained and present our conclusions and outlook for future studies. Finally, we report in Appendix a full account of the spectral technique and the notation adopted.

## 2. Data and Methods

**a. Reanalysis datasets**

We use the freely available southern hemisphere 500hPa daily geopotential height fields provided by the National Center for Environmental Prediction (NCEP dataset; Kistler et al., 2001) and by the European Center for Mid-Range Weather Forecast (ERA40 dataset; Simmons and Gibson, 2000) . We consider the datasets for the overlapping time frame ranging from September $1^{st}$ 1957 to August $31^{st}$ 2002. Both reanalyses are publicly released with spatial resolution of 2.5° x 2.5°, with a resulting horizontal grid of 144 x 73 points. We select the June-July-August (JJA) data averaged over the latitudinal belt 30°S-75°S (similarly to what done in D05 for the NH) , where the bulk of the baroclinic and of the low frequency planetary waves activity is observed (Simmonds and Keay 2000; Trenberth 1991).

The principal inhomogeneity in the assimilated data is the introduction of satellite observations (systematically used since 1979) which has dramatically improved the amount of available data over the last three decades (Sturaro, 2003). The first source of satellite data for reanalysis is the Vertical Temperature and Pressure Radiometer (VTPR), available from 1973 to 1978. However, whereas for the NCEP reanalyses raw satellite radiances at low spatial resolution are used to extract vertical temperature profiles, the ERA40 reanalysis assimilates directly the raw radiances at full resolution (See D05 for more details). Such a discrepancy related to the assimilation of satellite data, could be relevant in the SH where the ground based observing system is coarser than in the NH.

Moreover, in the SH the misplacing of PAOBS data in NCEP (Kistler et al 2001) for the period 1979-1992 is known to produce an error in the description of wintertime synoptic-scale features poleward of 40°S over the oceans that is comparable to the basic uncertainty of the analyses. (Kistler et al 2001, Simmonds and Keay 2000) .

In order to provide a rough estimate of the effect of this bug, we also consider the NCEP-DOE AMIP II reanalysis (hereafter NCEP2) that fixes this mistake and other previous processing errors present in NCEP (see Kanamitsu 2002, where a detailed description of NCEP2 is reported). NCEP2 has been released for the time frame 1979-2002 and should not be considered as a different reanalysis. Rather, it should be regarded as an updated and error–fixed version of NCEP. The physical parameterisations are the same used in NCEP, with the same spatial and temporal resolution, and the satellite data are treated in the same way.

**b.Spectral Analysis**

In a recent paper (D05), we have compared the winter midlatitude variability of the 500 hPa geopotential field of the NH using the space-time transformation introduced by Hayashi (1971). Such technique allows us to discriminate the variance associated with eastward/westward propagating waves and standing waves.

This information may be obtained by firstly Fourier-analysing the spatial field, and then computing the time-power spectrum of each spatial Fourier component. The difficulty here lies in the fact that straightforward space-time decomposition will not distinguish between standing and travelling waves: a standing wave will give two spectral peaks corresponding to travelling waves moving eastward and westward at the same speed and with the same phase. The problem can only be circumvented by making assumptions regarding the nature of the wave. For instance, we may assume complete coherence between the eastward and westward components of standing waves and attribute the incoherent part of the spectrum to real travelling waves (Pratt 1976; Fraedrich and Bottger 1978; Hayashi 1979).

In this formulation, for each winter considered, we express the 500hPa geopotential height $Z(\lambda,t)$, averaged on the latitudinal belt where the bulk of the variability occurs, in terms of the its zonal Fourier harmonic as:

(2.1) $\quad Z(\lambda,t) = Z_0(t) + \sum_{j=1}^{\infty} \{C_{k_j}(t)\cos(k_j\lambda) + S_{k_j}(t)\sin(k_j\lambda)\}$

where the zonal wavenumber is expressed as $k_j = 2j\pi$.

The power spectrum $H_{E/W}(k_j,\omega_m)$ at a zonal wavenumber $k_j$ and temporal frequency $\omega_m = 2\pi m/\tau$, where $\tau = 90d$ is the length of the winter, for the eastward and westward propagating waves is:

(2.2a) $\quad H_E(k_j,\omega_m) = \frac{1}{4}\{P_{\omega_m}(C_{k_j}) + P_{\omega_m}(S_{k_j})\} + \frac{1}{2}Q_{\omega_m}(C_{k_j}, S_{k_j})$

(2.2b) $\quad H_W(k_j,\omega_m) = \frac{1}{4}\{P_{\omega_m}(C_{k_j}) + P_{\omega_m}(S_{k_j})\} - \frac{1}{2}Q_{\omega_m}(C_{k_j}, S_{k_j})$

where $P_{\omega_m}$ and $Q_{\omega_m}$ are, respectively the power and the quadrature spectra of zonal Fourier harmonic of the 500hPa geopotential height $Z(\lambda,t)$.

The total variance spectrum $H_T(k_j,\omega_m)$ is given by the sum of the eastward and westward propagating components:

(2.3) $\quad H_T(k_j,\omega_m) = \frac{1}{2}\{P_{\omega_m}(C_{k_j}) + P_{\omega_m}(S_{k_j})\}$

while the propagating variance $H_P(k,\omega)$ is given by the absolute value of the difference between the components (2.2a) and (2.2b):

(2.4)    $H_P(k_j, \omega_m) = |Q(k_j, \omega_m)|.$

So, the standing variance spectrum $H_S(k,\omega)$ can be obtained by the difference:

(2.5)    $H_S(k_j, \omega_m) = H_T(k_j, \omega_m) - |Q(k_j, \omega_m)|.$

We emphasize that for sake of simplicity of the notation, we have neglected the indication of the winter under investigation, denoted in the text by the superscript *n*. We emphasize that customarily, Hayashi spectra are generally represented by plotting the quantities $j \cdot m \cdot H_T(k_j, \omega_m)$, $j \cdot m \cdot H_S(k_j, \omega_m)$, $j \cdot m \cdot H_E(k_j, \omega_m)$, and $j \cdot m \cdot H_W(k_j, \omega_m)$, in order for equal geometrical areas in the log-log plot to represent equal variance. With this definition, the Hayashi spectra presented in this paper are expressed in unit of $m^2$, as done in Blackmon (1976), Speranza (1983) and Lucarini et al. (2006), and can be compared to those given in D05 after a multiplication by *1/8\*86400s*

We have also introduced integral measures of the variance in order to characterize the different portion of the spectrum, and their temporal evolution. We introduce the following integral quantities:

(2.6)    $E_t^n(\Omega) = \sum_{m=m_1, j=j_1}^{m=m_2, j=j_2} H_j^n(k_j, \omega_m),$        with *t=T,S,E,W;*

where *n* indicates the winter; the integration extremes, $m_{1,2}$ and $j_{1,2}$, determine the spectral region of interest $\Omega = [\omega_{m_1}, \omega_{m_2}] \times [k_{j_1}, k_{j_2}]$. The quantity $E_t^n(\Omega)$ introduced in equation (2.6) represents the portion of variance of the spectrum associated to a given subdomain $\Omega$ and to a given winter *n* and is expressed in unit of $m^2$. The averaging process defined in equation (2.6) overcomes the well-known instability of the Fourier analysis in describing small scale spectral features. The quantities $E_t^n(\Omega)$ considered in this paper are more convenient than the corresponding integral quantities presented in D05. The $E_t^n(\Omega)$ can be compared numerically to the integral quantities presented in D05 after multiplying them by a multiplicative constant *1/8\*86400s*.

# 3. Hayashi spectra

## 3a. Climatological average

The Hayashi spectra express the power density of the wave field with respect to the frequency and the zonal wavenumber, and define the decomposition between the standing and propagating components. Fig. 1a shows $\overline{H}_T(k_j,\omega_m)$, the total power spectrum; Fig. 1b shows $\overline{H}_S(k_j,\omega_m)$, the power spectrum related to standing waves; Fig. 1c shows $\overline{H}_E(k_j,\omega_m)$, the power spectrum related to eastward propagating waves; Fig. 1d shows $\overline{H}_W(k_j,\omega_m)$, the spectrum of the westward propagating waves. The overbar indicates the operation of averaging over 45 JJA periods. Throughout the paper, $k_j = 2\pi j$ is the zonal wavenumber, $\omega_m = 2\pi m/\tau$ is the frequency of the wave with $\tau = 90d$ defined as length of each winter, and the indexes $T$, $S$, $E$, and $W$ indicate total, standing and eastward/westward propagating components, respectively.

The lobes of westward propagating components (Fig. 1d) are less intense and bounded on the planetary scale (zonal wavenumbers 2-3, periods longer than 20 days), as expected in a region where the large scale topography is present but small. The standing variance (Fig. 1b) is essentially comparable with that observed in the NH but has a more definite peak for k=3. This lobe of standing variance has been usually associated with blocking episodes (Trenberth and Mo 85, Raphael 2004). However, in the SH the total variability (Fig. 1a) is mainly explained by the eastward propagating component, which is more active than the corresponding one in the NH (see Fig. 1c in D05) and also extends to the planetary spatial scale for long periods (k=4 and period around 15 days), as illustrated in Fig. 1c. Note in particular that the amplitude of the variability associated to k=4 is twice as large as in the NH. The propagating component with zonal wavenumber 4 could be related to the Pacific/South America pattern PSA-1 and PSA-2 (Lau et al 1994). Recently, the real nature of these patterns of low-frequency variability has been debated in literature (Mo and Higgins 1998; Robertson and Mechoso 2003) Also in the spectral region of the baroclinic activity (zonal wavenumbers greater than 5 and periods less than 10 days) we observe a larger variance than in the NH. These results are in good agreement with the past analyses performed along the same lines but using shorter datasets (Fraedrich and Kietzig, 1983; Hansen et al.1989).

Following D05 we analyse the difference between the description of the waves climatology of the two reanalyses obtained by using the Hayashi technique. In Fig. 2 we report the difference between ERA40 and NCEP in the climatological spectra of the eastward propagating component. It clearly appears that ERA40 has a dramatically larger variance. The observed discrepancy corresponds to more than 25% of the signal. The other components do not show any relevant difference in the climatological averages. The discrepancies are most evident for zonal wave number k=5 and for a period of about 7 days, in the spectral region of baroclinic disturbances. The difference in the climatologies of the winter atmospheric variability in the SH is about one order of magnitude larger than what obtained in the NH. This result suggests that the two reanalyses reproduce very differently the process of baroclinic conversion of Available Potential Energy (APE). The agreement does not improve significantly when comparing ERA40 with NCEP2 for the overlapping period (1979-2002) (Figure not shown).

## 3b. Interannual variability

Following D05, we inspect the temporal behaviour of the previously observed discrepancies by using the integral quantities $E_t^n(\Omega)$, which represent the portion of variance associated to specific spectral subdomains $\Omega$ for each $n$ winter and with $t=T,S,E,W$. In the first two columns of Table 1 we report the time average $\overline{E}_t(\Omega)$ for the two reanalyses, computed over the whole wavenumber and frequency domain for the overlapping period 1957-2002. As in D05, we estimate the standard error of the time-averaged value with the interannual variability of the signal $\Delta_{E_t(\Omega)} = \frac{\sigma_{E_j(\Omega)}}{\sqrt{N}}$, where $N$ is the number of winters considered in the averaging process. This estimate is basically correct because, for our purposes, winters are statistically independent from each other. Most of the variance is due to the eastward propagating component, as can be inferred also from Fig. 1. The time-average of the ERA40 signal is considerably larger and the discrepancies between the averaged values for NCEP and ERA40 largely exceed the error bar related to the standard error $\Delta_{E_j(\Omega)}$, specially for the eastward propagating waves.

These features are confirmed when inspecting the time series of $E_t^n(\Omega)$ with $t=T,S,E,W$ for the two reanalyses (Fig. 3). The biases for the eastward propagating, and, less markedly, for the standing waves are apparent. In particular, Fig. 3 reveals significant discrepancies in the 1958-1972 time frame. There is an sudden jump in 1973, when the propagating eastward variance observed in ERA40 abruptly decreases with respect to NCEP. Another abrupt jump can be observed in 1978. Afterwards, ERA40 exhibits a systematic larger variance. Moreover, the ERA40 data feature a robust positive trend in the variability of the signal, especially for the eastward propagating waves, while in NCEP this property is less evident. However, recent studies stress the dubious correctness in the detection of ultra long term trends starting from reanalysis data (Bengtsson et al 2004).

Inspection of the differences in $E_t^n(\Omega)$ for the total, standing and eastward/westward propagating components (Fig. 4) displays more clearly the positive bias of ERA40. However, in the pre-satellite period the signal-to-noise ratio of the discrepancies between the two reanalyses is small.

Moreover, in the standing and westward propagating component there is no clear shift. This result does not change if we consider a different latitudinal belt (30S-50S or 50S-70S, not shown)
.

The abrupt change in the description of the variability corresponds to the onset of the assimilation of the VTPR data, as largely discussed in D05. This result suggests that the different use of satellite data (NCEP system assimilates retrieved profiles of temperature and humidity, ERA40 assimilates directly satellite radiance) corresponds to a very different capability in describing atmospheric variability, principally the travelling disturbances and the baroclinic conversion processes. This effect is particularly evident in the southern hemisphere where a weaker constraint on the models is provided by the sparse available land-based vertical soundings.

In order to compare homogeneously derived data, we skip the VTPR period 1973-1978. We show in Table 1 averaged values $\overline{E}_t(\Omega)$ for the pre and post-satellite periods, 1958-1972 and 1979-2002, respectively. In both periods the ERA40 mean values are clearly larger than for NCEP with differences significantly greater than the corresponding standard error $\Delta_{E_j(\Omega)}$. The abrupt change in $\overline{E}_{T,E}(\Omega)$ between the first and the second period in ERA40 is evident, as already pointed out by Bengtsson at al (2003), who nevertheless analysed different climatological variables. In NCEP the time averaged values seem to be more homogeneous. Nevertheless, a positive trend in the variance can be recognized by comparing the time averages above of two periods, as also noted by Renwick and Revell (1999)

In order to analyse the effects on the description of the midlatitude variability of the bugs found out in the original version of the NCEP data, we also compare the differences in $E_t^n(\Omega)$ for NCEP and NCEP2 for the overlapping period 1979-2002  As shown in Fig. 5, the discrepancies between the two releases of the NCEP-NCAR reanalysis are one order magnitude smaller than those found between NCEP and ERA40 and comparable to the bias between NCEP and ERA40 in the NH (D05), so they cannot be considered as negligible. In particular, the discrepancies are more pronounced in the first years, when NCEP data are affected by the well known PAOBS bugs, and all the components $E_t^n(\Omega)$ features similar biases. Nevertheless, we conclude that the known bugs in the NCEP reanalysis have a relatively minor role in the observed discrepancy between the NCEP/NCAR and ERA40 reanalyses.

Following D05, in Table 2 we propose the clear-cut division of different spectral sub-domains $\Omega$ into four categories, on the basis of the results reported in Fig. 1a-d. This partition can help identifying discrepancies in the capability of the two reanalyses in describing phenomena occurring on a given spatial and temporal scale. As in the NH, in the austral winter most of the variance has to be attributed to the Low Frequency- Low Wavenumber (LFLW) and High Frequency High Wavenumber (HFHW) waves (values not shown). However, the division is slightly different by that adopted for the NH for the differing spectral properties of the ultra long and synoptic waves in the southern region, above described.

In Fig. 6a-b we plot the time series of the difference between the quantities $E_t^n(\Omega_{LFLW})$ and $E_t^n(\Omega_{HFHW})$ for the two reanalyses. The qualitative properties shown in Fig. 4 are substantially confirmed. Regarding the discrepancy for the LFLW domain (Fig. 6a), the signal-to-noise ratio is small in the pre-satellite period, while, after a sudden jump during the VTPR period, the differences become systematic. ERA40 has, on average, a smaller standing variance than NCEP. In the HFHW spectral subdomain (Fig. 6b) the discrepancies are noisy in the earlier period, but ERA40 has systematically more variance than NCEP. After 1979, the signal-to-noise ratio becomes larger and the bias is almost constant.

# 4. Spatial distribution of variability before 1973 and after 1979

In Figs. 7a-d, we map the low frequency (LF) variability of NCEP for the periods 1958-72 (Fig. 7a) and 1979-2002 (Fig. 7b), and the corresponding differences with the same data for the ERA40 reanalysis (Fig. 7c-7d, respectively). The LF atmospheric variability is obtained by discarding the frequencies $\omega > 2\pi/11\ day^{-1}$ (see also Table 2) in the Fourier transform of the 500hPa geopotential height. The two maps of the NCEP variance show that the LF atmospheric variability has a more zonal character than in the NH (D05), even though two maxima over the South Pacific and, secondarily, over the South Indian ocean are well defined. In particular, in the South Pacific, the lobe associated to the PSA-1/-2 pattern is clearly evident (Trenberth and Mo 1985, Lau et al. 1994, Robertson and Mechoso 2003). South Pacific is the region where the major discrepancies between the two reanalyses can be observed. In particular, before 1973 (Fig. 7a) discrepancies in the location and intensity of blocking episodes and PSA are evident. This suggests a zonal shift between NCEP and ERA40 in the maximum of the LF variability. A similar shift, but oriented in the North-South direction, is also present over the Atlantic basin. In the Indian sector, ERA40 overestimates the LF variance with respect to NCEP. South of Australia, the differences are almost comparable with the signal itself. After 1979, ERA40 has systematically more variance than NCEP, especially over south Pacific region but no spatial shift occurs anymore.

The spatial pattern of the high frequency (HF) variability ($10\ day^{-1} \leq \omega \leq 2\pi/2\ day^{-1}$, as in Table 2) is reported in Figs. 8a-d. In Fig. 8a-b the well known pattern of the southern storm track is shown. The storm track is characterised by a greater zonal symmetry than in the NH and peaks around 50°S over the Indian Ocean (Trenberth 1991). As reported in Fig. 8c-d, in both periods, ERA40 has more HF variability than NCEP on the entire southern mid-latitude band; the relative differences are of the order of 30%, with peaks of about 40% over the Indian Ocean. Considering that the meridional heat transport has a maximum over this region (Pexioto and Oort, 1992), our hypothesis on the discrepancies between the two reanalyses in the Lorenz energy cycle could be confirmed. However, unlikely what observed in Fig. 7 for the LF, in the HF band the introduction of the satellite does not drastically modify the discrepancies observed between the two reanalyses. Therefore, we can guess that satellite data do not constrain sufficiently the models for having the same behaviour in the description of the baroclinic processes.

## 5. Discussion and conclusions

In this paper the NCEP and ERA40 reanalyses have been compared for the overlapping period 1957-2002. Differences in the description of the southern hemisphere winter variability have been quantitatively estimated by employing the space-time Fourier decomposition introduced by Hayashi (1971; 1979) and recently re-discussed in D05. On average, ERA40 is characterised by a variance larger than NCEP, mainly in the region of the eastward propagating planetary (k=4, period≅15d) and long synoptic waves (k=5; period≅7d).

By using *ad hoc* integral measures, we have inspected the temporal behaviour of the southern hemisphere variability in the two reanalyses. In particular we find a sudden jump in the ERA40 variability in the '70s, in agreement with what reported by Bengtsson et al. (2004). We have found that the estimates given by the two reanalyses for both high frequency and low frequency disturbances are roughly unbiased with random discrepancies in the pre-satellite period (especially for LFLW), while abrupt jumps in the biases are apparent in the period corresponding to the introduction of VTPR data (1973-1978), and eventually a systematic positive bias for ERA40 in the satellite period is apparent. The differences are very large, up to about 30% of the signal. No better accordance with ERA40 has been found using NCEP2 data, where some known bugs present in NCEP have been fixed.

Complementarily, we have inspected the differences between the two reanalyses in their representation of low and high frequency variability at geographical level by suitably filtering the geopotential height field and estimating its variance at each grid point. In the low frequency window, we observe geographical shifts in the maximum of the wave activity over the Pacific and Atlantic Oceans, while, after 1979, the discrepancies are systematic, with ERA40 featuring a larger variance. In the high frequency window, ERA40 features larger variances throughout the considered time-frame, with an especially large bias in the southern storm track region in the Indian Ocean during the post-satellite period.

From these results some conclusion can be pointed out:

1. The discrepancies between the two reanalyses in the SH are in general of the order of 10-20% and are about one order of magnitude larger than what reported in the NH in D05: this implies that today we do not have a well-defined picture of the statistical properties of the winter atmospheric variability of the mid-latitudes of the SH to be used in the evaluation of the realism of climate models.

2. In the pre-satellite period a weaker constraint on the models is provided by the sparse land-based vertical soundings available. In this period, the disagreement of the background field dominates. Even after the introduction of satellite data in the assimilation systems, differences in the models and the different treatment of the data lead to a different representation of variability.

3. The differences observed in the LF modes can be ascribed to the position and intensity of SH blocking episodes and recurrent spatial patterns PSA-1/2 (among others, Renwick and Revell 1999) and could be related to a different representation of the tropical forcing in the two reanalyses. This discrepancy is probably due to a well known differences in the representation of tropical OLR, in particular over the deadline (Mo and Higgins, 1998) . However, the nature of LF variability in the South Pacific is greatly debated in literature and a deeper analysis is out of the aim of this work. In the HF spectral region the systematic bias observed suggests that the operational models improved for the two reanalyses follow a different behaviour in the baroclinic conversion of Available Potential Energy (APE), as also suggested in D05.

4. Some caveat could be introduced regarding to what asserted by Kidson (1999) about the period 1970-73 when a different behaviour of atmospheric variability is observed. During this period different data and different assimilation methods have been used in the assimilation systems. Thus, a caution should be used for direct comparison of atmospheric fields in different periods. The recently observed trends in the southern hemisphere could be influenced by the major changes in the observing system which occurred in '70s (see also Bengtsson et al 2004)

Finally, we wish to emphasize that this work provides the motivation, the methodologies, and the diagnostics tools to be employed for the analysis of the degree of self-consistency (realism, as said before, cannot be assessed) of climate models simulations, *e.g.* those provided in the context of Program for Climate Model Diagnosis and Intercomparison (PCMDI), sponsored by the Intergovernmental Panel on Climate Change Assessment Report 4 (IPCC-AR4), along the lines of Lucarini et al. (2006).

Acknowledgments.
This research was partly supported by the Italian National Programme of Antarctic Research (P.N.R.A.). V.L. wishes to thanks ISAC-CNR for the kind hospitality. The authors wish to thank Antonio Speranza and Alfonso Sutera for useful discussions..

# References

Bengtsson, L., S. Hagemann, and K. I. Hodges (2004) Can climate trends be calculated from reanalysis data?, *J. Geophys.Res.*, 109, D11111, doi:10.1029/2004JD004536.

Benzi R, Malguzzi P, Speranza A Sutera A (1986) The statistical properties of general atmospheric circulation: observational evidence and a minimal theory of bimodality. *Q J Roy Met Soc* 112: 661-674

Benzi R, Speranza A (1989) Statistical properties of low frequency variability in the Northern Hemisphere. *J Clim* 2: 367-379

Blackmon ML (1976) A climatological spectral study of the 500 mb geopotential height of the Northern Hemisphere. *J Atmos Sci* 33: 1607-1623

Bromwich DH, Fogt RL (2004) Strong trends in the skill of the ERA40 and NCEP/NCAR reanalyses in the high and middle latitudes of the southern hemisphere, 1958–2001. J Clim

Cai, M. and Mak, M. (1990) Symbiotic relation between planetary and synoptic scale waves. *J. Atmos. Sci.* 47 2953-2968.

Charney, J.G. (1947) The dynamics of long waves in a baroclinic westerly current *J. Meteor.*, 4, 135-162.

Charney JG, DeVore JG (1979) Multiple flow equilibria in the atmosphere and blocking. *J Atmos Sci* 36: 1205-1216

Charney JG, Straus DM (1980) Form-drag instability, multiple equilibria and propagating planetary waves in the baroclinic, orographically forced, planetary wave system. J Atmos Sci 37: 1157-1176

Cuff T J, Cai M. (1995) Interaction between the low and high-frequency transients in the southern hemisphere winter circulation. Tellus 47A, 331-350

Dell'Aquila, A., Lucarini, V., Ruti, P.M., and S. Calmanti (2005) Hayashi Spectra of the Northern Hemisphere Mid-latitude Atmospheric Variability in the NCEP-NCAR and ECMWF Reanalyses. *Climate Dynamics,* DOI: 10.1007/s00382-005-0048-x.

Eady, E.T.(1949) Long waves and cyclone waves. *Tellus,*1, 35-52.

Fraedrich K, and Bottger H (1978) A wavenumber frequency analysis of the 500 mb geopotential at 50°N. *J Atmos Sci* 35: 745-750

Fraedrich, K. and E. Kietzig, (1983) Statistical analysis and wavenumber frequency spectra of the 500 mb geopotential along 50°S. *J. Atmos. Sci.* 40, 1037-1045.

Frederiksen, J. S., and C. S. Frederiksen (1993) Southern Hemisphere storm tracks, blocking, and low-frequency anomalies in a primitive equation model. J. Atmos. Sci., 50, 3148–3163

Hansen, A.R., Sutera, A. and D.E. Venne (1989) An examination of midlatitude power spectra : evidence for standing variance and the signature of El Nino. *Tellus*, 41A, 371-384.

Hayashi.Y.(1971) A generalized method for resolving disturbances into progressive and retrogressive waves by space Fourier and time cross-spectral analysis. *J. Meteorol. Soc. Jap.*,49, 125-128.

Hayashi Y. (1979) A generalized method for resolving transient disturbances into standing and travelling waves by space-time spectral analysis. *J. Atmos. Sci.,* 36, 1017-1029.

Holton JR (1992) An introduction to dynamic meteorology. Academic Press Boston, 497 pp

Hoskins, B.J., and K.I. Hodge (2002): New Perspectives on the Northern Hemisphere Winter Storm Tracks. *J. Atmos. Sci.,* **59**, 1041-1061.

James, I. N. (1994) Introduction to Circulating Atmospheres. Cambridge University Press, 246 pp.

Kanamitsu M., Ebisuzaki E., Woollen J. , Shi-Keng Yang, J. J. Hnilo, M. Fiorino, and G. L. Potter (2002) NCEP–DOE AMIP-II Reanalysis (R-2). *Bull. Am. Meteor. Soc., 83*, 1631-1643.

Kidson, J. W. (1999) Principal modes of southern hemisphere low-frequency variability obtained from NCEP-NCAR reanalysis. *J. Climate,* 12*,* 2808-2830.

Kistler R, Kalnay E, Collins W, Saha S, White G, Woollen J, Chelliah M, Ebisuzaki W, Kanamitsu M, Kousky V, van den Dool H, Jenne R, Fiorino M, (2001) The NCEP-NCAR 50-year reanalysis: Monthly means CD-ROM and documentation. *Bull. Am. Meteorol. Soc.* 82, 247–267

Lau, K.-M., P.-J. Sheu, and I.-S. Kang, (1994) Multiscale low-frequency circulation modes in the global atmosphere. *J. Atmos. Sci.,* 51, 1169–1193.

Lucarini V., S. Calmanti, A. Dell'Aquila, P.M. Ruti, and A. Speranza, (2006) Intercomparison of the northern hemisphere winter mid-latitude atmospheric variability of the IPCC models, submitted to Climate Dynamics, ArXiV preprint: http://arxiv.org/abs/physics/0601117

Mechoso, C.R., and D.L. Hartmann, (1982) An observational study of travelling planetary waves in the Southern Hemisphere. *J. Atmos. Sci.* 39, 1921-1935,

Mo, K. C. and R. W. Higgins (1998) The Pacific South American modes and tropical convection during the Southern Hemisphere winter. *Mon. Wea. Rev.,* 126, 1581–1598.

Peixoto, J. P. and Oort, A. H. (1992) *Physics of Climate.* American Institute of Physics, New York, 520 pp.

Pedlosky, J. (1979) *Geophysical Fluid Dynamics.* Springer-Verlag, New York, 624 pp.

Pratt RW (1976) The interpretation of space-time spectral quantities. *J Atmos Sci* 33: 1060–1066

Renwick, J.A. and M. J. Revell, (1999) Blocking over the South Pacific and Rossby Wave propagation. *Mon. Wea. Rev.* 127, 2234-2247.

Robertson, A. W., and C. R. Mechoso (2003): Circulation Regimes and Low-Frequency Oscillations in the South Pacific Sector. Mon. Wea. Rev., 131, 1566-1576

Ruti, P. M., V. Lucarini, A. Dell'Aquila, S. Calmanti, and A. Speranza (2006), Does the subtropical jet catalyze the midlatitude atmospheric regimes?, Geophys. Res. Lett., 33, L06814, doi:10.1029/2005GL024620.

Simmonds, I.and K. Keay, (2000) Mean Southern Hemisphere extratropical cyclone behavior in the 40-year NCEP–NCAR reanalysis. *J. Climate,* 13, 550-561

Simmons, A. J. and J. K. Gibson (2000) The ERA-40 Project Plan, *ERA-40 Project Report Series No. 1, ECMWF*, 62 pp.

Speranza A (1983) Determinist and statistical properties of the westerlies. *Paleogeophysics* 121: 511-562

Sterl A., 2004: On the (in-)homogeneity of reanalysis products, *J. Climate*., 17, 3866-3873

Sturaro, G., (2003): A closer look at the climatological discontinuities present in the NCEP/NCAR reanalysis temperature due to the introduction of satellite data. *Climate Dyn.,* 21, doi:10.1007/s00382-003-0348-y.

Tibaldi, S., E. Tosi, A. Navarra, and L. Peduli, (1994): Northern and Southern Hemisphere seasonal variability of blocking frequency and predictability. *Mon. Wea. Rev.* 122, 1971-2003.

Trenberth, K E, (1991): Storm track in the southern hemisphere. *J. Atmos. Sci.* 48, 2159-2178.

Trenberth, K.E. and Mo, K.C. (1985). Blocking in the southern hemisphere. *Mon. Wea. Rev.* 113, 3-21.

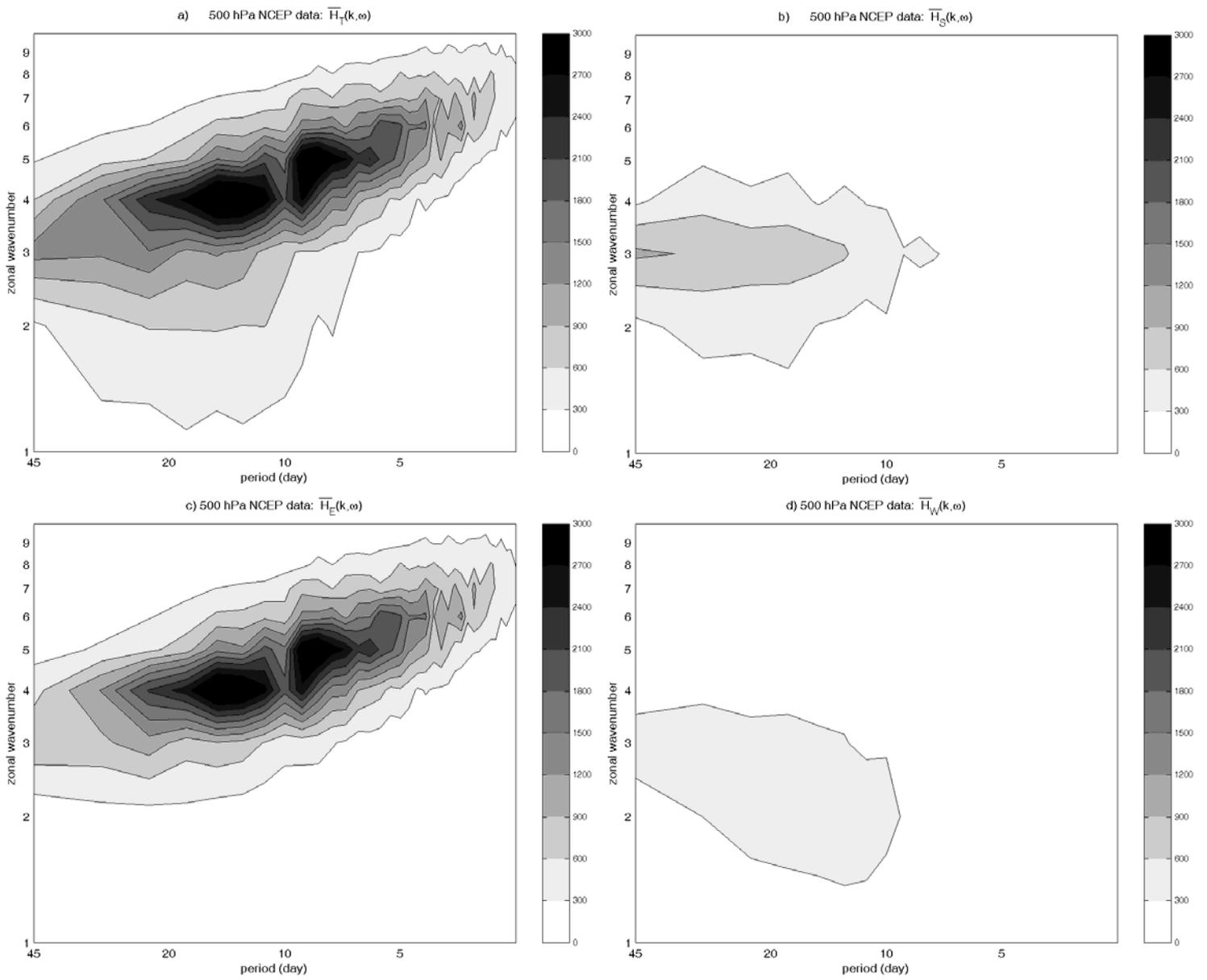

**Figure 1**: Climatological average over 45 winters of Hayashi spectra for 500 hPa geopotential height (relative to the latitudinal belt 30°S-75°S) from NCEP data: $\overline{H}_T(k_j, \omega_m)$ (a); $\overline{H}_S(k_j, \omega_m)$ (b); $\overline{H}_E(k, \omega)$ (c); $\overline{H}_W(k_j, \omega_m)$ (d). The Hayashi spectra have been obtained multiplying the spectra by *j\*m*. The units are $m^2$

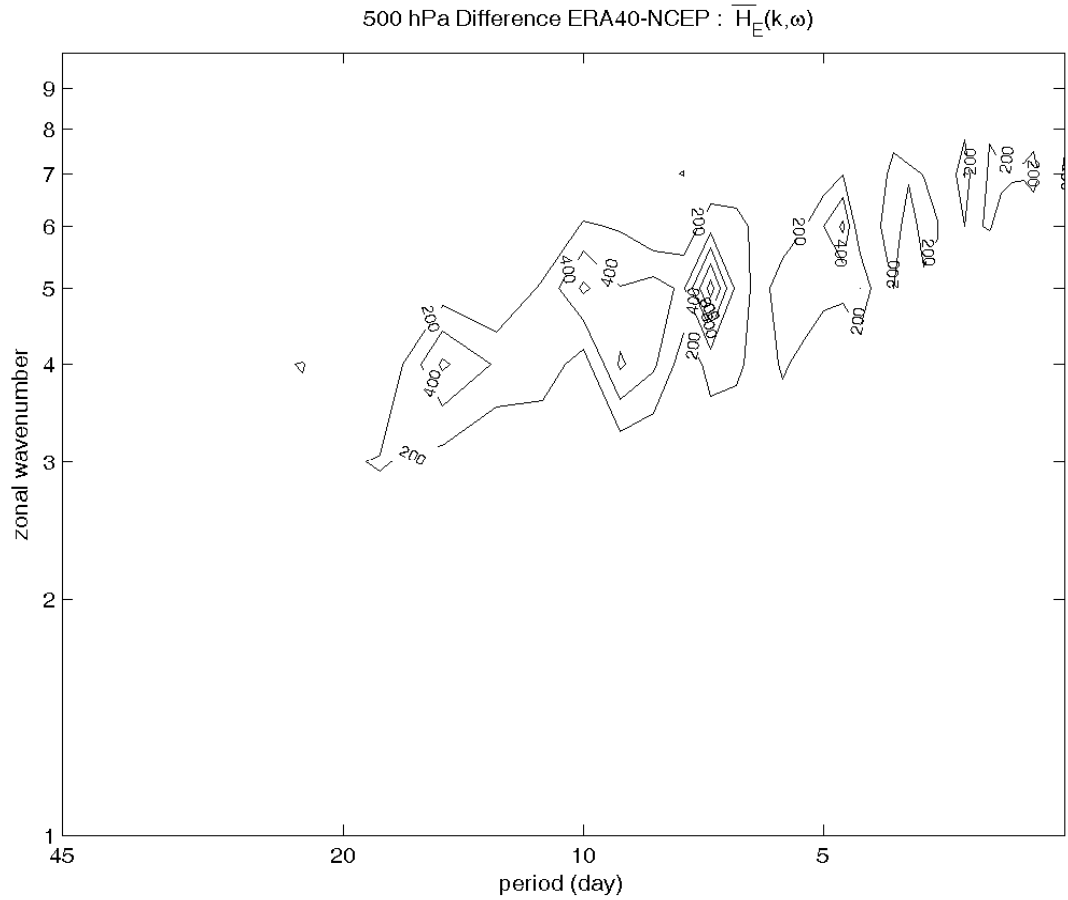

**Figure 2:** As in Fig.1c but for the difference between ERA40 and NCEP dataset Hayashi spectra. We do not plot the zero contours. In the standing and westward propagating component there are no significant differences and are not reported here.

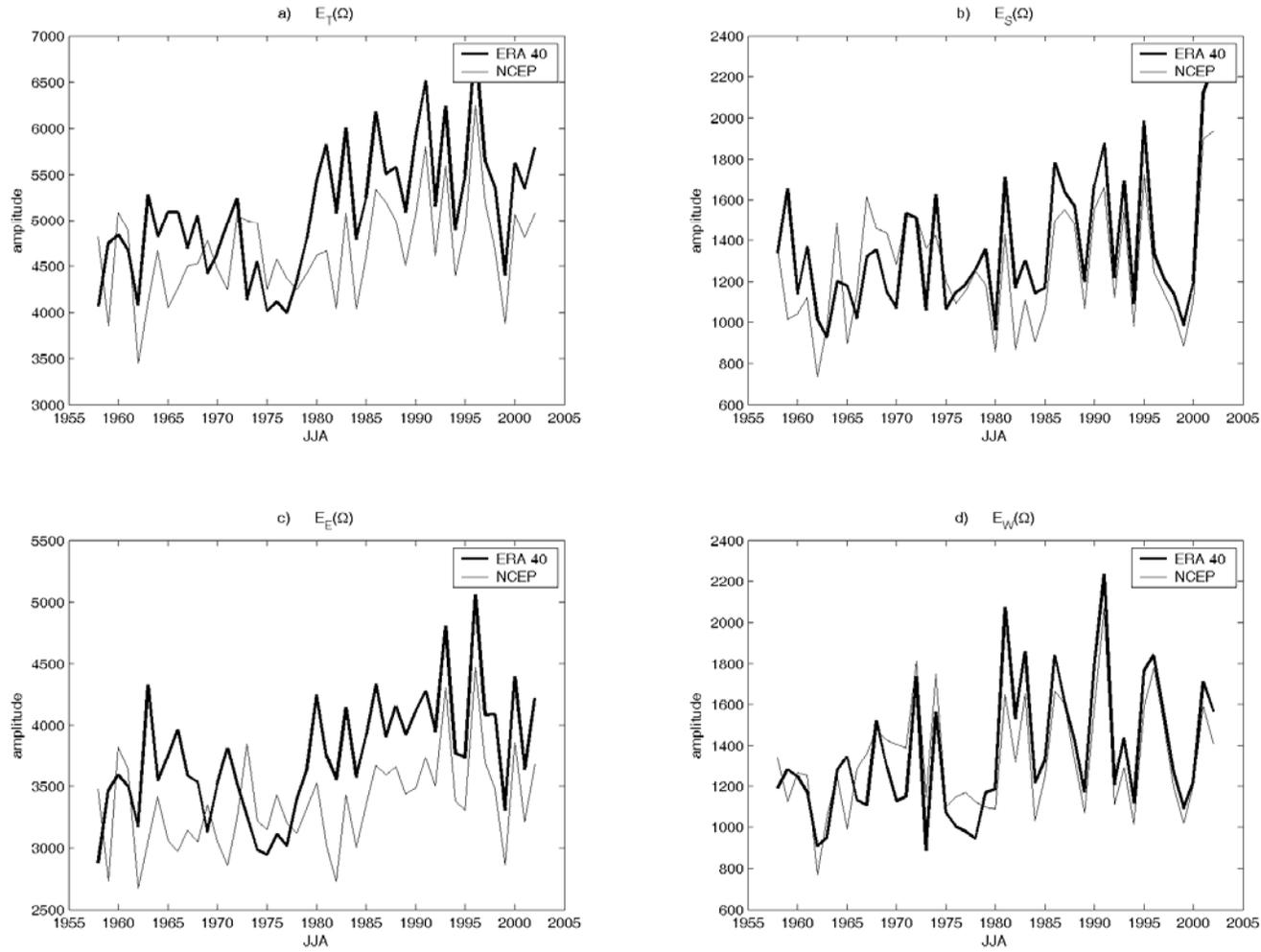

**Figure 3**: Time series of the quantities $E_j^n(\Omega)$ computed for the NCEP (thin line) and ERA40 (thick line) for total variance - a) panel - , standing – b) panel -, propagating eastward – c) panel -, and propagating westward – d) panel. Here, $\Omega$ corresponds to the whole wave number and frequency domain.

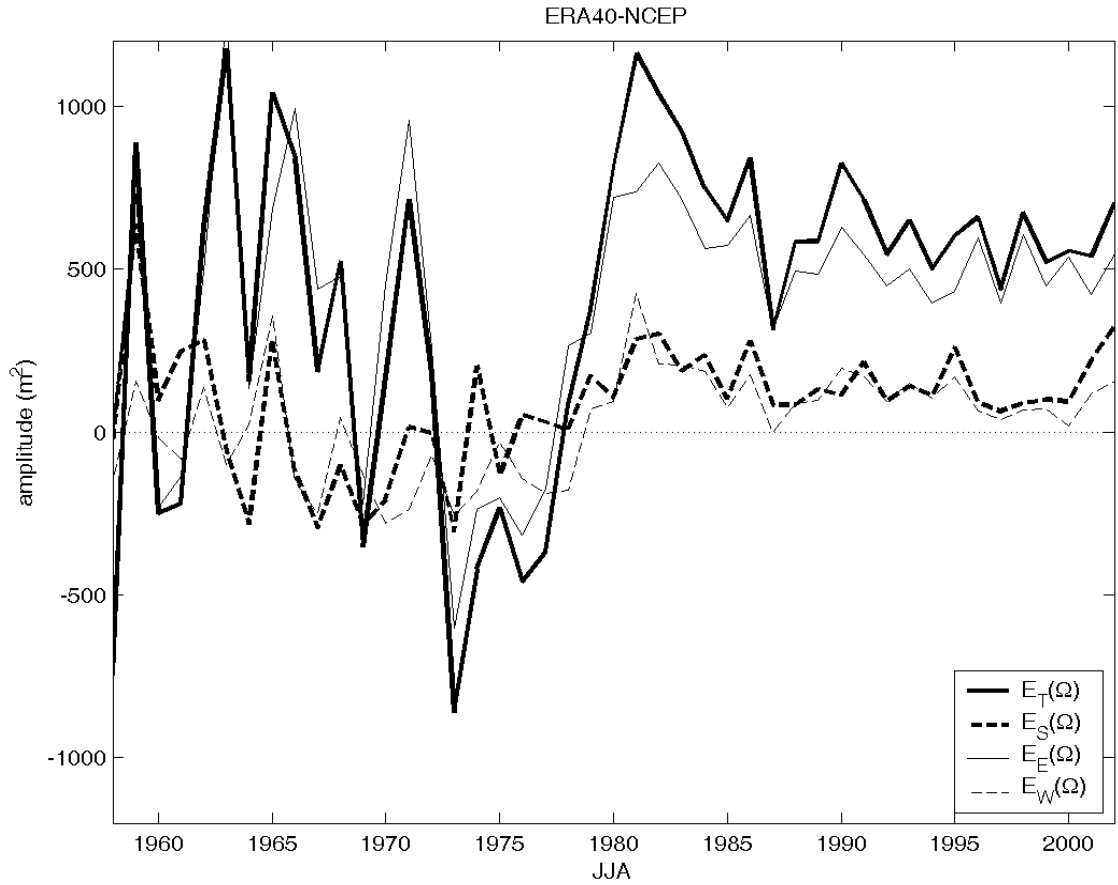

**Figure 4:** Time series of the differences of the quantity $E_j^n(\Omega)$ computed for the two reanalysis datasets, for: total variance (thick line), standing (dashed thick line), propagating eastward (thin line) and propagating westward (dashed thin line). Here, $\Omega$ corresponds to the whole wave number and frequency domain.

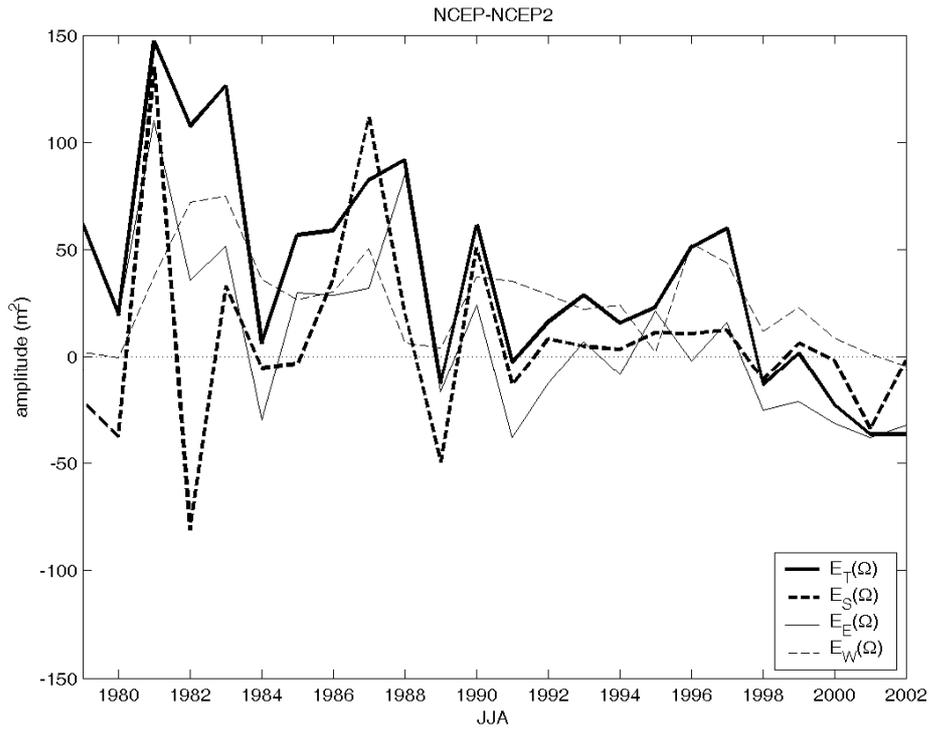

**Figure 5:** As in figure 3 for the differences between NCEP and NCEP2 .

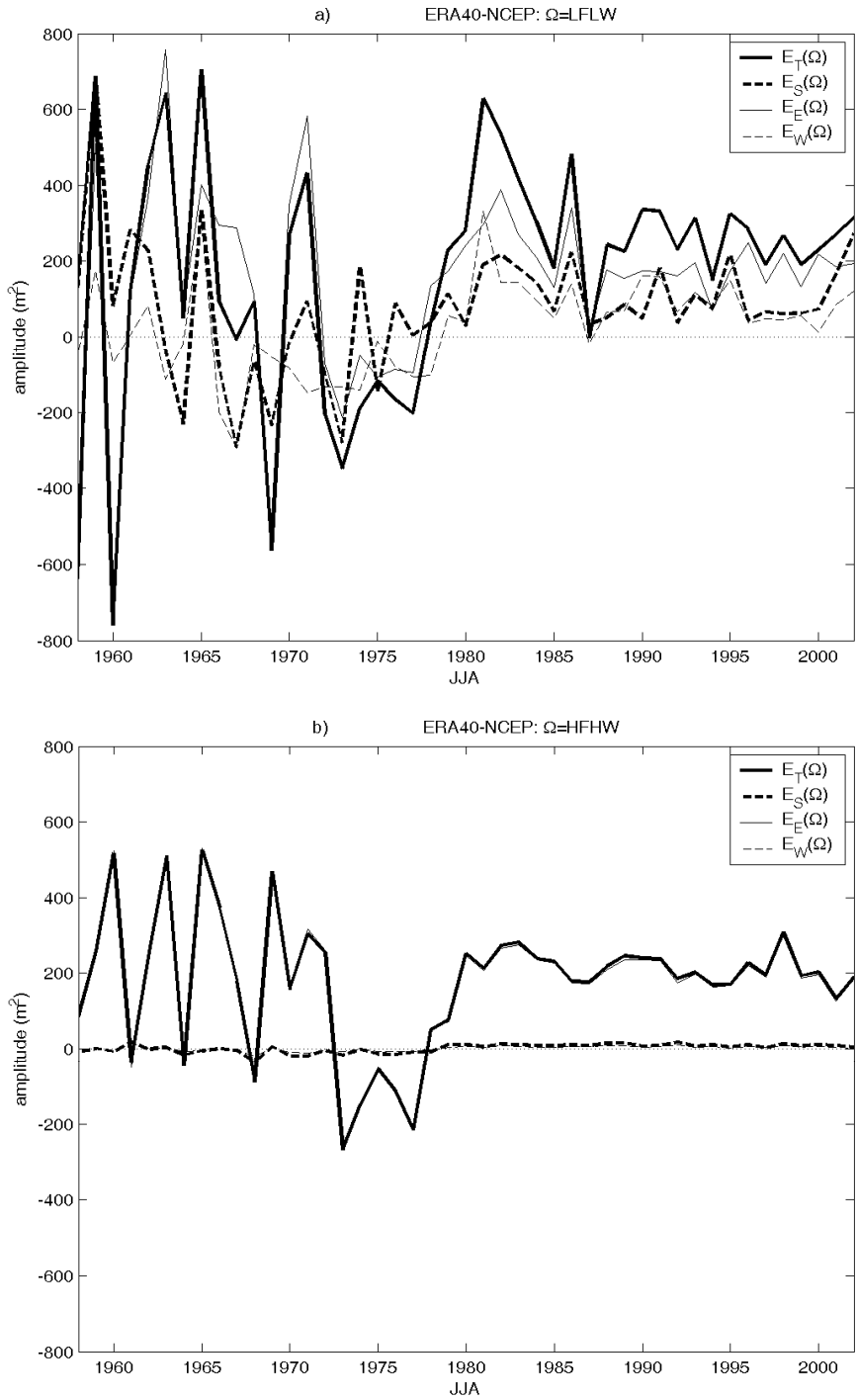

**Figure 6:** As in figure 3 for the categories *LFLW (a), HFHW (b)* described in Table 2.

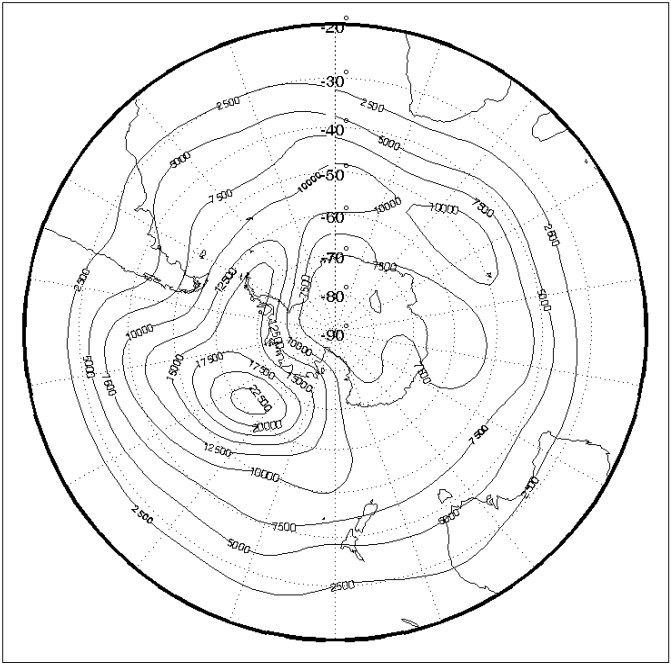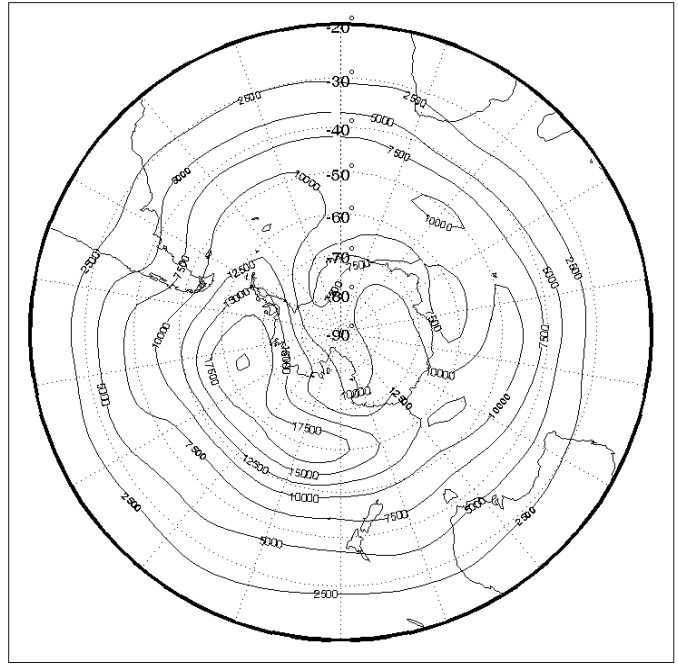
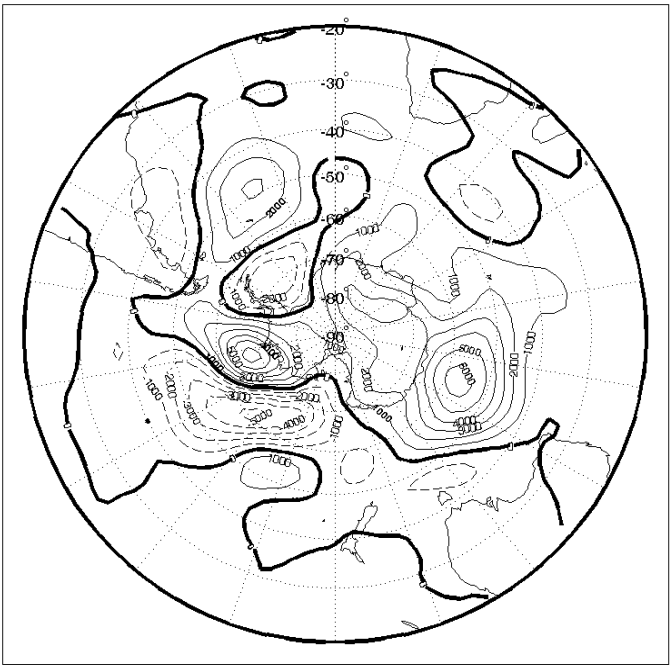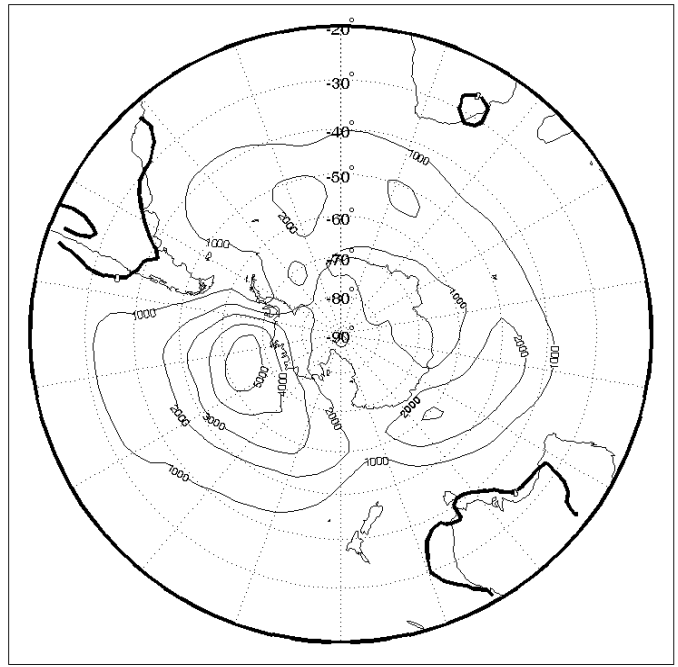

**Figure 7:** Low frequency variance of the 500 hPa geopotential height, JJA period: a) NCEP 1958-1972; b) NCEP 1979-2002; c) difference between ERA40 and NCEP, 1958-72; d) as c) for 1979-2002. Units are $m^2$ and contour interval is 2500 $m^2$ for a) and b) and 1000 $m^2$ for c) and d). Dashed lines pertain to negative values. Thick line corresponds to zero difference.

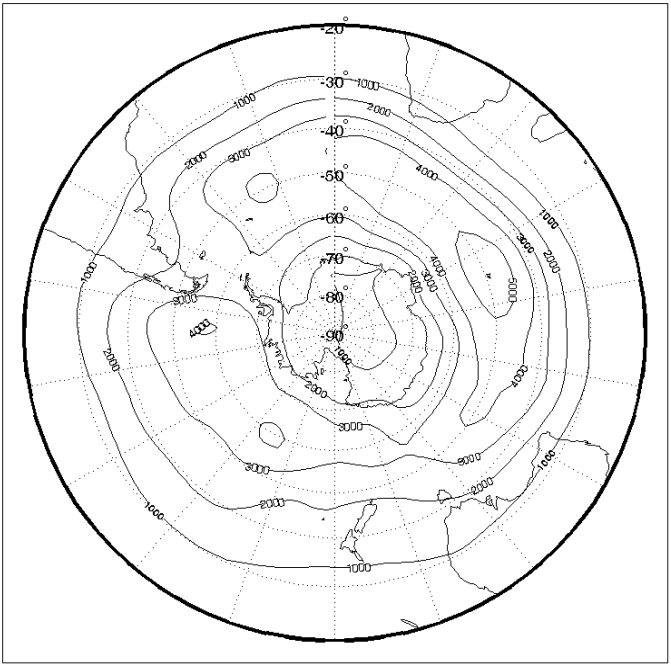
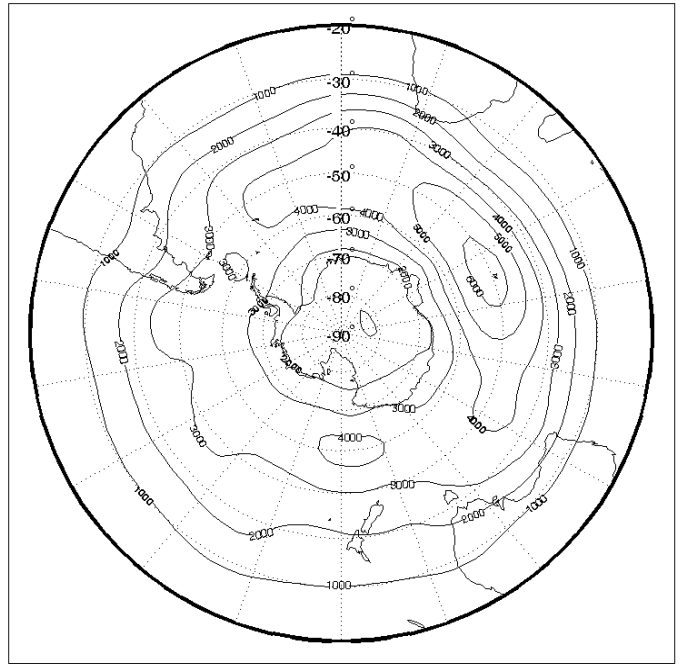
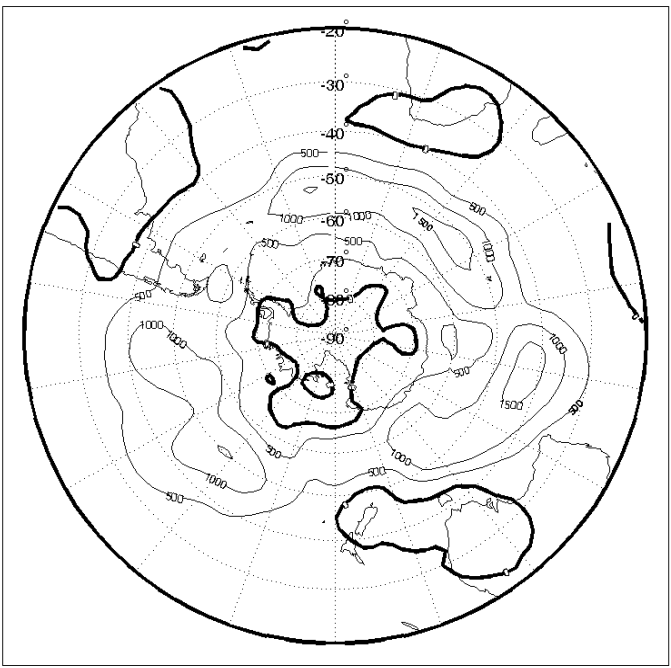
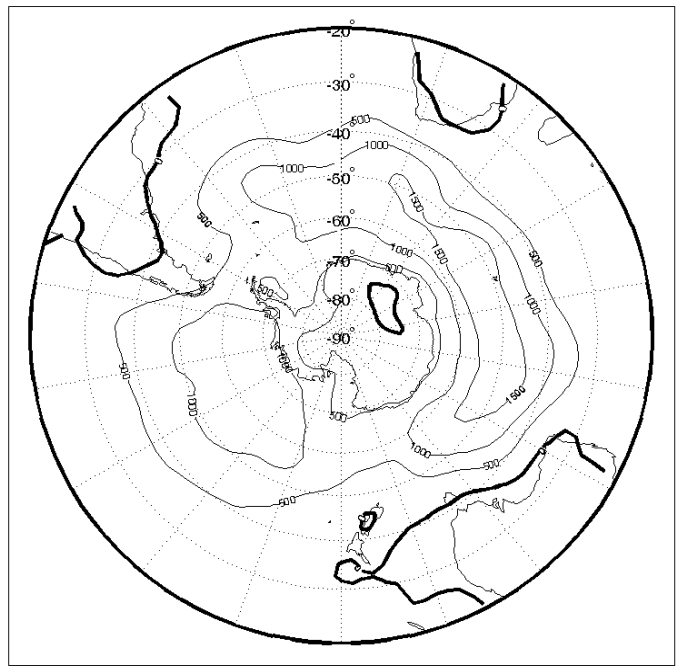

**Figure 8:** High frequency variance of the 500 hPa geopotential height, JJA period: a) NCEP 1958-1972; b) NCEP 1979-2002; c) difference between ERA40 and NCEP, 1958-72; d) as c) for 1979-2002. Units are $m^2$ and contour interval is 1000 $m^2$ for a) and b) and 500 $m^2$ for c) and d). Dashed lines pertain to negative values. Thick line corresponds to zero difference.

TABLES

| Ω=ALL | ERA 1957-2002 | NCEP 1957-2002 | ERA 40 1957-72 | NCEP 1957-72 | ERA 40 1979-2002 | NCEP 1979-2002 |
|---|---|---|---|---|---|---|
| $\overline{E}_T(\Omega)$ | 5100 ±100 | 4690±70 | 4610±90 | 4480±90 | 5600±100 | 4900±100 |
| $\overline{E}_S(\Omega)$ | 1350±50 | 1270±40 | 1250±40 | 1240±50 | 1430±70 | 1230±70 |
| $\overline{E}_E(\Omega)$ | 3740±70 | 3360±60 | 3430±80 | 3220±70 | 4040±90 | 3490±80 |
| $\overline{E}_W(\Omega)$ | 1360±40 | 1330±40 | 1180±50 | 1260±50 | 1520±70 | 1340±60 |

**Table 1:** Time mean of $E_j(\Omega)$ with *j=T, S, E, W* for the JJA period of the whole record, 1957-1972 and 1979-2002, respectively. We consider the standard error of the time-averaged value as a function of the interannual variability of the signal: $\Delta_{E_j(\Omega)} = \frac{\sigma_{E_j(\Omega)}}{\sqrt{N}}$. As in Fig.3, $\Omega$ corresponds to the whole wave number and frequency domain. The units are $m^2$

| Spectral properties | $k_1 = 2, k_2 = 4$ | $k_1 = 5, k_2 = 72$ |
|---|---|---|
| $\omega_1 = (2\pi/45)d^{-1}, \omega_2 = (2\pi/11)d^{-1}$ | $\Omega$ = LFLW | $\Omega$ = LFHW |
| $\omega_1 = (2\pi/10)d^{-1}, \omega_2 = (2\pi/2)d^{-1}$ | $\Omega$ = HFLW | $\Omega$ = HFHW |

**Table 2:** Definition of 4 regions in the Hayashi spectra of the winter atmospheric variability; the symbol $d$ is used as shorthand for 'day'. *LFLW*: Low Frequency Long Wavenumber; *HFLW*: High Frequency Long Wavenumber; *LFHW*: Low Frequency High Wavenumber; *HFHW*: High Frequency High Wavenumber. Low Frequency relates to periods from 11 to 45 days; High Frequency relates to periods from 2 to 10 days; Low Wavenumber relates to length scales larger than 1000Km; High wavenumber relates to length scales ranging from a few tens to hundreds of kilometres. The values $\omega_2 = (2\pi/2)d^{-1}, k_2 = 72$ constitute the highest frequency and wavenumber allowed by the adopted data resolution.